\documentclass[lettersize,journal]{IEEEtran}
\usepackage{amsmath,amsfonts}
\usepackage{array}
\usepackage{textcomp}
\usepackage{stfloats}
\usepackage{url}
\usepackage{verbatim}
\usepackage{graphicx}
\usepackage{cite}
\usepackage{color}
\graphicspath{{./figures/}{./graphics/}{./graphics/logos/}}
\usepackage{svg}
\usepackage{lscape}
\usepackage{subcaption}
\usepackage{multirow}
\usepackage{threeparttable}
\usepackage{booktabs}
\usepackage{colortbl}
\usepackage{psfrag}
\usepackage{amsmath}
\usepackage{amssymb}
\usepackage{amsfonts}
\usepackage{amsthm}
\usepackage{mathrsfs}
\usepackage{breqn}
\usepackage{braket}

\usepackage{algorithm}
\usepackage{algorithmicx}
\usepackage{algpseudocode}
\usepackage{paralist}
\usepackage{url}
\usepackage{hyperref}
\usepackage{cleveref}
\usepackage{multicol}
\usepackage[english]{babel}
\usepackage{blindtext}
\usepackage{algorithm}
\usepackage{algorithmicx}
\usepackage{algpseudocode}
\usepackage{paralist}
\usepackage{subcaption}
\usepackage{amsmath}
\usepackage{url} \IEEEoverridecommandlockouts
\usepackage{enumitem} 
\usepackage{pifont}

\providecommand{\DIFdel}[1]{} 

\usepackage{tcolorbox}
\usepackage{xcolor}
\usepackage{listings}
\usepackage{tabularx} 
\usepackage{graphicx}
\usepackage{xcolor}

\definecolor{codegreen}{rgb}{0,0.6,0}
\definecolor{codegray}{rgb}{0.5,0.5,0.5}
\definecolor{codepurple}{rgb}{0.58,0,0.82}
\definecolor{backcolour}{rgb}{0.95,0.95,0.92}

\lstdefinestyle{mystyle}{
    backgroundcolor=\color{backcolour},   
    commentstyle=\color{codegreen},
    keywordstyle=\color{magenta},
    numberstyle=\tiny\color{codegray},
    stringstyle=\color{codepurple},
    basicstyle=\ttfamily\footnotesize,
    breakatwhitespace=false,         
    breaklines=true,                 
    captionpos=b,                    
    keepspaces=true,                 
    numbers=left,                    
    numbersep=5pt,                  
    showspaces=false,                
    showstringspaces=false,
    showtabs=false,                  
    tabsize=2
}

\begin{document}
\title{sat-QFL: Secure Quantum Federated Learning for Low Orbit Satellites}
\author{Dev Gurung and Shiva Raj Pokhrel, Senior Member, IEEE 
\thanks{Authors are with the QUANTIMA Research, School of IT, Deakin University, Geelong, Australia}
\thanks{Manuscript received April 19, 2025; revised August 16, 2026.}}
\markboth{Journal of \LaTeX\ Class Files,~Vol.~14, No.~8, August~2026}%
{Shell \MakeLowercase{\textit{et al.}}: A Sample Article Using IEEEtran.cls for IEEE Journals}
\maketitle

\begin{abstract}
Low Earth orbit (LEO) constellations violate core assumptions of standard (quantum) federated learning (FL): client-server connectivity is intermittent, participation is time varying, and latency budgets are strict. We present sat-QFL, a hierarchical, access aware quantum federated learning (QFL) framework that partitions satellites into primary (ground connected) and secondary as inter-satellite links (ISL-only) roles, and schedules sequential, simultaneous, or asynchronous edge training aligned with visibility windows. For quantum-resilient confidentiality and integrity, sat-QFL integrates quantum key distribution (QKD) based key establishment with authenticated encryption for model exchange; we also assess teleportation as a feasibility primitive for quantum state transfer. Using derived constellation traces and QFL workloads (Qiskit), we show that sat-QFL sustains robust aggregation under varying participation and reduces communication bottlenecks with modest security overhead. Our implementation and results are available at \url{https://github.com/s222416822/satQFL}.

\end{abstract}

\section{Introduction}
Recent work has advanced federated learning (FL) for satellite communications (SatCom)~\cite{10418548}, targeting collaborative on-orbit training across low Earth orbit (LEO) constellations \cite{elmahallawyCommunicationEfficientFederatedLearning2024, wangFederatedLearningLEO2022a, wuDSFLDecentralizedSatellite2022}. LEO systems already support large-scale Earth observation and SatCom, generating substantial imagery and sensor data \cite{yangCommunicationEfficientSatelliteGroundFederated2024}, which enable time-critical applications including disaster response \cite{chenSatelliteBasedComputingNetworks2022}, meteorology \cite{wuFedGSMEfficientFederated2023}, and broadband connectivity \cite{ 
maralSatelliteCommunicationsSystems2009}.

SatCom extends coverage to remote regions \cite{wangNodeSelectionStrategy2024} amid rapid IoT growth,
yet faces intermittent access, dynamic line-of-sight (LoS), and stringent latency/energy constraints. Consequently, SatCom is a key complement to terrestrial networks for AI services \cite{chenSatelliteBasedComputingNetworks2022}, and a central enabler in 6G NTN roadmaps where LEO constellations are prominent \cite{chenSatelliteBasedComputingNetworks2022}.

\begin{figure}
    \centering
    \includegraphics[width=0.9\linewidth]{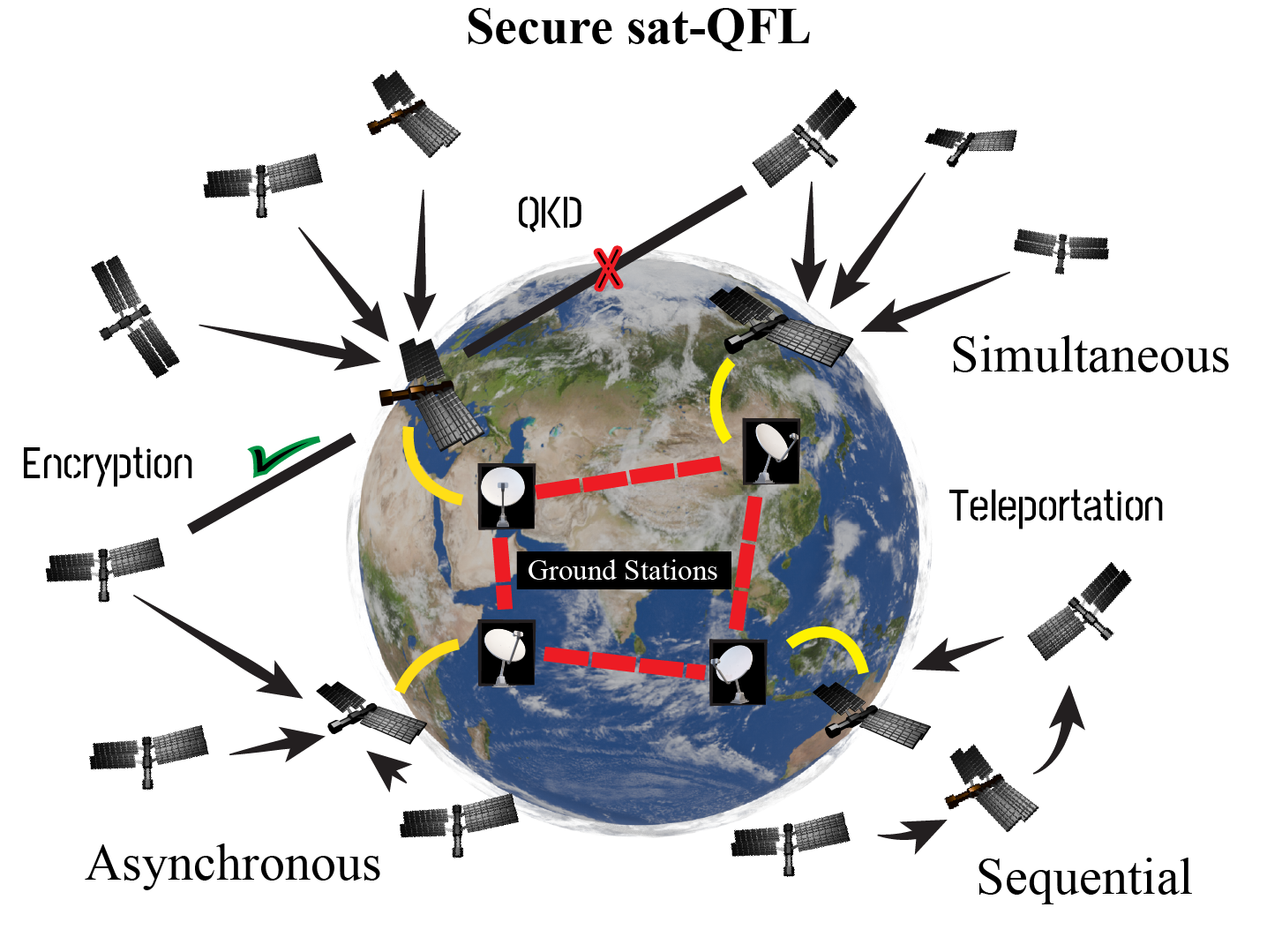}
    \caption{sat-QFL preview: Various frameworks for sat-QFL proposed in this work all secured with quantum key distribution along with classical cryptography; ({\color{green}\checkmark} Access; {\color{red}\ding{55}} No Access) }
    \label{fig:satQFL_preview}
\end{figure}
FL, introduced by McMahan et al. \cite{mcmahanCommunicationEfficientLearningDeep2023}, trains models collaboratively without sharing raw data, reducing backhaul and improving privacy. Quantum federated learning (QFL) extends these benefits to quantum settings \cite{gurung_chained_2025, chen_introduction_2024}, enabling quantum clients to train locally on quantum data with server-side aggregation \cite{gurung_performance_2024, chen_introduction_2024}. Adapting QFL to satellite networks follows the trajectory of classical SatFL \cite{soFedSpaceEfficientFederated2022, wuClientSelectionSatellite2024, wuDSFLDecentralizedSatellite2022, chenSatelliteBasedComputingNetworks2022}, but must address constellation-specific constraints: server dependence, convergence under partial participation, and bidirectional bottlenecks \cite{pokhrelBlockchainBringsTrust2021a}. Training can span days to weeks \cite{elmahallawyCommunicationEfficientFederatedLearning2024}, and reliable links require LoS.

Security is equally critical and satellite links can leak model parameters to eavesdroppers \cite{williamsExplorationsQuantumComputing2011}. Beyond classical cryptography, 
quantum cryptography, particularly QKD (Quantum Key Distribution), which offers information-theoretic key establishment grounded in quantum mechanics \cite{bennett_quantum_2014}.

Building on aforementioned observations, we introduce a \textit{novel QFL framework tailored to LEO constellations}, refer to as \textbf{sat-QFL}; a high-level design details of which is shown in Figure~\ref{fig:satQFL_preview}. 
To conclude
our modeling and the design of sat-QFL in this research have three key contributions:\\
(i) a topology and access aware design for LEO that jointly accounts for orbital dynamics and QFL constraints;\\
(ii) a security stack by design that combines QKD and teleportation with cryptography for confidential and authenticated model exchange; and\\
(iii) theoretical and empirical analyses that demonstrate practicality and robustness in LEO's settings.
\begin{table*}[t]
\centering
\caption{Comparison of related FL/QFL works for satellite networks}
\label{tab:related_works}
\scriptsize
\setlength{\tabcolsep}{3pt}
\begin{tabularx}{\linewidth}{l|c|c|c|c|c}
\toprule
Work&Scope&Topology&Aggregation&Scheduling&Security/Quantum \\
\midrule
Chen \cite{chenSatelliteBasedComputingNetworks2022} & LEO+Gnd & Partial & Centralized & Sync/Light & Classical / -- \\
AsyncFLEO \cite{elmahallawyAsyncFLEOAsynchronousFederated2022} & LEO & 
Ring-of-stars
& Centralized & Async & Classical / -- \\
NomaFedHAP \cite{elmahallawyCommunicationEfficientFederatedLearning2024} & LEO+HAP & HAP-aided & Dist. PS
& Window-aware & Classical / -- \\
DSFL \cite{wuDSFLDecentralizedSatellite2022} & LEO & 
Energy-Aware
& Decentralized & 
Intra/Inter Plane
& Classical / -- \\
SatelliteFL \cite{yangCommunicationEfficientSatelliteGroundFederated2024} & Sat–Gnd & Explicit windows & Centralized & Window-opt. & Classical / -- \\
EdgeSel/Cluster \cite{chenEdgeSelectionClustering2023} & LEO & Cluster-aware & Hierarchical & Selected & Classical / -- \\
FedHAP \cite{elmahallawyFedHAPFastFederated2022} & LEO+HAP & Coverage-aware & Collab HAP & 
Synchronous
& Classical / -- \\
FedLEO \cite{zhaiFedLEOOffloadingAssistedDecentralized2024} & LEO & Offload-aware & Decentralized & 
On-orbit offloading
& Classical / -- \\
Ground/On-Board \cite{razmiGroundAssistedFederatedLearning2022}
& LEO & Gnd vis. & On-board/GA & 
Async
& Classical / -- \\
FedGSM \cite{wuFedGSMEfficientFederated2023} & Sat–Gnd & 
Time-varying
& Centralized &
Async
& Classical / -- \\
FedSecure
\cite{elmahallawySecureEfficientFederated2023} & LEO & 
Partial
& On-orbit+DKG 
& Partial & Classical DKG / -- \\
OliveBranch \cite{fangOliveBranchLearning2021} & SAGIN & 
Ring/Two-Tier Star
& Hierarchical & 
Configurable
& Classical / -- \\
FedSL \cite{abediFedSLFederatedSplit2022,jiangFederatedSplitLearning2024} & STIN & N/A & Split & Sequential Data & Classical / -- \\
LEOShot
\cite{elmahallawyOneShotFederatedLearning2023}
& LEO & 
Multi-Orbit/Sat
& Hybrid & One-shot & Classical / -- \\
SFL-LEO \cite{hassanSFLLEOSecureFederated2023} & 6G LEO & 
NTN
& 
N/A
& 
Network Operator
& Classical / -- \\
Blockchain FL \cite{pokhrelBlockchainBringsTrust2021a} & UAV/IoT($\rightarrow$Sat) & Limited & Decentralized trust & Async & Blockchain / -- \\
Wang \cite{wangFederatedLearningLEO2022a} & Gnd–LEO & Limited & LEO server & Sync & Classical / -- \\
QFL frameworks \cite{chen_introduction_2024,gurung_performance_2024,gurung_chained_2025} & General & N/A & Centralized/Sequential & N/A & -- / Quantum \\
\midrule
\textbf{Secure sat-QFL (ours)} & ISL+Gnd & LoS/windows & Hierarchical  & Seq/Sim/Async (window-aligned) & QKD+AEAD;Tel/Quantum\\
\bottomrule
\end{tabularx}
\end{table*}
\subsection{Related Work}
Prior work has made tangible progress on \emph{classical} federated learning (FL) over LEO constellations~\cite{10418548}, but it does not resolve the fundamental mismatches that arise when directly porting (quantum) FL to dynamic satellite topologies. Chen et al. \cite{chenSatelliteBasedComputingNetworks2022} outlined SatCom-oriented FL to reduce backhaul and enable adaptive learning at scale, while AsyncFLEO \cite{elmahallawyAsyncFLEOAsynchronousFederated2022} mitigates synchronous idle time via asynchronous approach.
Using high-altitude platforms (HAP) as distributed parameter servers (PS) (NomaFedHAP) \cite{elmahallawyCommunicationEfficientFederatedLearning2024} improves satellite visibility
and blockchain-based coordination \cite{pokhrelBlockchainBringsTrust2021a} targets trust and auditability. SatelliteFL \cite{yangCommunicationEfficientSatelliteGroundFederated2024} optimizes per-window training/utility under intermittent ground access. Decentralized schemes \cite{wuDSFLDecentralizedSatellite2022} and LEO-centric edge selection and clustering \cite{chenEdgeSelectionClustering2023} further reduce dependence on terrestrial servers.

A comprehensive comparison of related FL/QFL advances for satellite is shown in Table~\ref{tab:related_works}. Despite these relevant advances including
split learning \cite{jiangFederatedSplitLearning2024,abediFedSLFederatedSplit2022}, HAP collaboration \cite{elmahallawyFedHAPFastFederated2022}, offloading-assisted decentralization \cite{zhaiFedLEOOffloadingAssistedDecentralized2024}, ground-assisted scheduling \cite{razmiGroundAssistedFederatedLearning2022}, 
one-shot 
propagation 
\cite{elmahallawyOneShotFederatedLearning2023}, 
resource optimization \cite{jingResourceOptimizationSignal2021}, and secure/efficient on-orbit schemes \cite{elmahallawySecureEfficientFederated2023, hassanSFLLEOSecureFederated2023},
the prevailing assumptions remain \emph{classical}: (i) model representations and training dynamics are non-quantum; (ii) security relies on classical cryptography; and (iii) system designs often presume either a reachable server or sufficiently stable participation sets per round. These assumptions break down for QFL and, more generally, for learning over LEO constellations with time-varying LoS, intermittent ground access, strict latency/energy budgets, and ISL constraints.

With QFL~\cite{pokhrel2024quantum}, first it introduces quantum data, parameterizations, and aggregation that are sensitive to delayed, missing, or out-of-date updates; classical asynchronous fixes do not directly translate to quantum ansatz and measurement-driven gradients. Second, classical security is not sufficient against quantum-capable adversaries~\cite{pokhrel2024poster}; QFL deployments require quantum-resilient key establishment (e.g., QKD) integrated with authenticated encryption, largely absent from prior SatFL frameworks~\cite{gurung_performance_2024}. Third, dynamic topology and access-window scheduling must be \emph{first-class} in the learning protocol: uniform client participation and synchronous server availability are unrealistic in LEO, and ad-hoc remedies (relays, offloading) do not provide a general, topology-aware learning abstraction compatible with quantum workloads.

In wireless resource optimization, Narottama et. al \cite{NarottamaQNNParallelTrainingWireless} has proposed parallel training based on Quantum neural network where wireless resource optimization is done through statistical information from edge devices.
Some works have successfully achieved quantum teleportation of independent single-photon qubits from a ground to a LEO satellite through uplink channel covering distance of 1,400 km \cite{renGroundtosatelliteQuantumTeleportation2017}.
Similarly, Lioa et. al \cite{liaoSatellitetogroundQuantumKey2017}
developed and launched LEO satellite for decoy-state QKD achieving kilohertz key rate over a distance of 1,200 km paving the way for a global scale quantum networks.
Nevertheless, directly applying standard QFL~\cite{pokhrel2024quantum} to LEO is inadequate. Our primary objective in this paper is to develop a hierarchical, access-aware QFL framework that (a) explicitly partitions satellites into primary (ground-connected) and secondary (ISL-only) roles for aggregation, (b) schedules sequential/simultaneous/asynchronous edge training aligned with visibility windows, and (c) integrates QKD-assisted keying with classical authenticated encryption to achieve quantum-resilient confidentiality and 
integrity-addressing 
the unresolved gaps in the above literature.

\subsection{Problem Formulation}
\begin{figure}[!htb]
    \begin{subfigure}[b]{\columnwidth}
        \centering
        \includegraphics[width=\columnwidth]{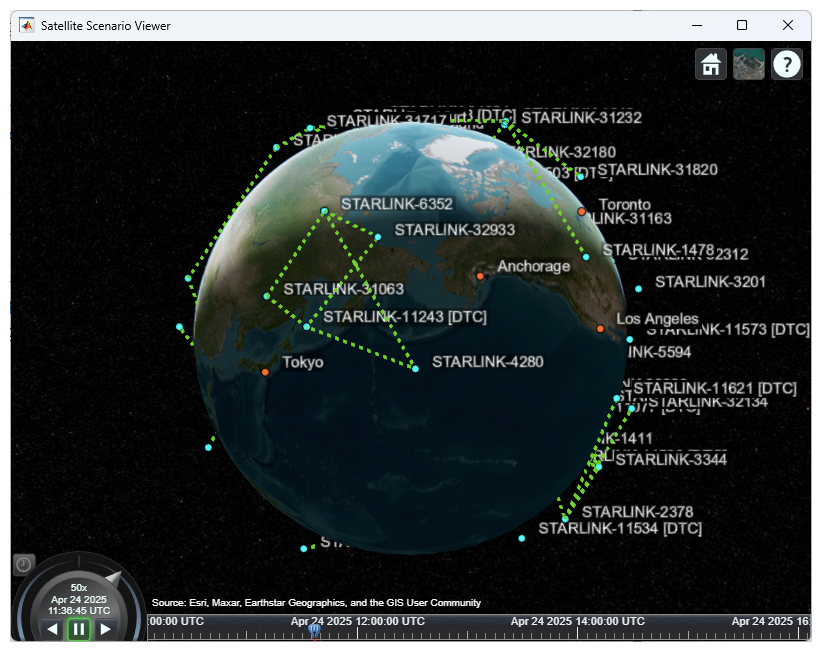}
    \end{subfigure}
    \caption{Problem Formulation: Satellite constellations vary that can impact the implementation of federated learning framework into this kind of networks.}
    \label{fig:constellationn}
\end{figure}

Figure~\ref{fig:constellationn} illustrates a Starlink snapshot where time-varying line-of-sight (LoS) induces dynamic connectivity clusters among satellites. At any given instant, only a subset of satellites has ground-station visibility (e.g., Tokyo may reach STARLINK-31063 and STARLINK-11243[DTC]), while others rely solely on inter-satellite links (ISLs). These properties violate standard FL/QFL assumptions of stable, simultaneous client-server connectivity, hindering synchronous aggregation and uniform participation.

To model this realistically, let $n$ denote the number of satellites. We partition them into: (i) \emph{primary} satellites with current ground access and (ii) \emph{secondary} satellites without ground access that can forward updates via ISLs to primaries. This primary-secondary abstraction captures the constellation’s time-varying topology and motivates topology and access-aware QFL protocols for practical deployment in LEO constellations.

Beyond learning protocols, practical satellite deployments face power/link budgets, transceiver design, and end-to-end link simulation. Here, we focus on a federated learning protocol tailored to LEO constellations (e.g., Starlink), leaving physical-layer co-design to future work.

Consider $n$ LEO satellites $\mathcal{S}=\{s_1,\dots,s_n\}$ and $m$ ground stations $\mathcal{G}=\{g_1,\dots,g_m\}$. Let the time-varying connectivity be an undirected graph $\mathcal{H}(t)=(\mathcal{V},\mathcal{E}(t))$, $\mathcal{V}=\mathcal{S}\cup\mathcal{G}$, where edges represent line-of-sight (LoS) at time $t\in[t_0,t_f]$. An inter-satellite edge $(s_i,s_j)\in\mathcal{E}(t)$ exists if their positions $\mathbf{p}_i(t),\mathbf{p}_j(t)\in\mathbb{R}^3$ satisfy a LoS criterion (e.g., view angle $\le\theta_{\max}$). The induced subgraph over $\mathcal{S}$ yields time-varying connected components (LoS clusters), complicating synchronous FL/QFL.

Define the set of \emph{primary} satellites (ground-visible at $t$)
\[
\mathcal{S}_p(t)=\{\,s\in\mathcal{S}\mid \exists g\in\mathcal{G}:\ (s,g)\in\mathcal{E}(t)\,\},
\]
and \emph{secondary} satellites $\mathcal{S}_s(t)=\mathcal{S}\setminus\mathcal{S}_p(t)$, which can only reach ground via inter-satellite links (ISLs) to $\mathcal{S}_p(t)$. Let $\mathcal{P}_i(t)\in\{0,1\}$ indicate the existence of a feasible path from $s_i$ to some $g\in\mathcal{G}$ in $\mathcal{H}(t)$ subject to constraints (e.g., hop count $H\le H_{\max}$, latency $L\le L_{\max}$). The participating set at time $t$ is
\[
\mathcal{C}(t)=\{\,i\in\{1,\dots,n\}\mid \mathcal{P}_i(t)=1\,\}.
\]

Each satellite $s_i$ holds data $\mathcal{D}_i$ and model parameters $\theta_i(t)$. Local updates follow
\[
\theta_i(t{+}1)=\theta_i(t)-\eta\,\nabla f(\theta_i(t),\mathcal{D}_i),
\]
and are transmitted only if $\mathcal{P}_i(t)=1$. Ground aggregation at time $t$ uses received updates:
\[
\theta_g(t{+}1)=\frac{1}{|\mathcal{C}(t)|}\sum_{i\in\mathcal{C}(t)} \theta_i(t{+}1).
\]
The global learning objective is
\[
\min_{\theta}\ \frac{1}{n}\sum_{i=1}^n \mathbb{E}_{(x,y)\sim\mathcal{D}_i}\big[f(\theta; x,y)\big],
\]
subject to $\mathcal{H}(t)$-induced participation and routing constraints. In typical snapshots (e.g., 50-satellite scenario), only a subset (e.g., 22) is ground-visible while the rest (e.g., 28) require ISLs, breaking uniform participation and synchronous rounds.

We therefore propose an access-aware QFL framework that (i) prioritizes primaries for direct aggregation, (ii) routes secondary updates via ISLs under $(H_{\max},L_{\max})$, and (iii) schedules edge training (sequential/simultaneous/asynchronous) to align with access windows, thereby improving robustness under partial connectivity while mitigating latency and compression/quantization errors. In fact, we have made the following three considerations for designing pragmatic QFL for satellites.

\textbf{Assumption 1 (Bounded staleness).} \textit{There exists $\Delta_{\max}$ such that any local update used at time $t$ was computed at time $t'\ge t-\Delta_{\max}$.}

\textbf{Assumption 2 (Participation).} \textit{For each window $W$, the probability a satellite participates is at least $p_{\min}>0$; access windows are ergodic over $[t_0,t_f]$.}

\textbf{Proposition 1 (Convergence under partial participation).} \textit{Under Assumptions 1–2 and standard smoothness/PL or convexity conditions for the QML loss, sat-QFL with weighted aggregation and step-size $\eta_t\propto 1/\sqrt{t}$ converges in expectation to a neighborhood whose radius scales with $\Delta_{\max}$ and $(1-p_{\min})$. }








\section{Preliminaries and Background}
\subsection{Quantum Federated Learning (QFL)}
QFL~\cite{gurung_chained_2025, pokhrel2024quantum} combines quantum machine learning
with federated learning to take advantage of both paradigms. 
In a typical QFL setup, multiple clients \(c_i\) with quantum processors collaboratively train a global
quantum model \(Q(\theta)\) while keeping their local data \(\delta_i\) private. 
Each client \(c_i\) performs local updates on its quantum model parameters \(\theta_i\) using a local loss 
function \(f_i(\theta_i)\), 
and then shares the updated parameters \(\theta_i\) with a central server. 
The server aggregates these parameters to update the global model \(\theta\) using an aggregation rule such as  \(\theta = \frac{1}{N}\sum_{i=1}^N w_i \theta_i\), where \(w_i\) represents the weighting factor for the client \(c_i\). This iterative process can be formulated as
$\theta_i^{t+1} = \theta_i^{t} - \eta \nabla_{\theta_i} f_i(\theta_i^{t})$
where \(t\) denotes the iteration step, \(\eta\) is the learning rate, and \(\nabla_{\theta_i} f_i(\theta_i^{t})\) is the loss function gradient with respect to local parameters.

\subsection{Satellite Communications}

Satellite communications transmit and receive radio-frequency (RF) signals between spaceborne relays and ground terminals to interconnect distant locations. At microwave/mmWave bands, propagation is largely line-of-sight (LoS) and obstructed by Earth’s curvature, motivating spaceborne transponders that receive, amplify/convert, and forward signals between geographically separated endpoints \cite{10418548}.

Low Earth orbit (LEO) satellites (altitude $\sim$160 - 2{,}000\,km) offer lower latency, reduced path loss, and lower energy per bit compared to higher orbits \cite{chenSatelliteBasedComputingNetworks2022}. However, LEO systems face challenges including intermittent ground visibility, limited feeder-link bandwidth, handovers, Doppler, and susceptibility to channel outages or attacks \cite{yangCommunicationEfficientSatelliteGroundFederated2024}.

Communication satellites typically operate in three orbital regimes: geostationary Earth orbit (GEO), medium Earth orbit (MEO), and LEO. Recent advances have enabled large LEO constellations with dense inter-satellite and feeder links; their shorter propagation delays make them attractive for time-sensitive services (e.g., interactive broadband, edge AI) relative to MEO/GEO.



\subsection{Quantum Key Distribution}
Quantum cryptography is different from conventional cryptography in that the data is kept secret on the basis of quantum mechanics \cite{shor_simple_2000}.
QKD is a secure communication scheme that is based on the principles of quantum physics used for symmetric cryptographic keys \cite{zahidy_practical_2024}.
One of the first proposed QKD protocols was called BB84 \cite{bennett_quantum_2014}.
In this protocol, the participating parties share a secret key through a quantum channel through series of qubits sent through various rotation bases while the receiver follows the protocol to retrieve the key information. 
In brief, sender generates random bits of sequences 0's and 1's. 
From those bits, the sender randomly selects a base and prepares quantum states.
At receiver, for qubits sent by sender through a quantum channel, random bases are selected, and the qubits are measured. 
The receiver is unaware of what base was used by the sender. 
The receiver then maintains the record of the measurement results and the associated base used.
Based on the statistics, receiver is supposed to choose right bases half the time.
Using a classical channel, the sender informs of the base that was used to encode the key.
By comparison, bits obtained measuring qubits in different bases are discarded (key sifting).
In Algorithm \ref{alg:qkd}, 
The QKD protocol is presented which follows the standard BB84 \cite{brassardTeleportationQuantumComputation1998a} protocol.
The main idea is that both sender and receiver with pre-entangled qubit pair share secret key securely.
The key is based on the size of message to be encrypted, etc.
Also, the use of base states to rotate the qubits states and matching which bases are used by both sender and receiver further helps in creating keys.

\subsection{Teleportation}
Quantum teleportation enables the transmission of quantum information (quantum state) to a distant location, presented in Algorithm \ref{alg:teleportation} \cite{brassardTeleportationQuantumComputation1998a}.
For teleportation, the two parties involved need to share the prior quantum entanglement.
To implement quantum teleportation, we need a quantum circuit containing certain gates.
With quantum teleportation, the so-called ``Quantum Internet" is believed to be possible.
As based on quantum mechanics principle, teleportation relies on both classical and quantum channel (entanglement) between two communication parties \cite{liaoSatellitetogroundQuantumKey2017}.
In quantum teleportation, we have two parties, sender and receiver. 
The scenario is that sender wants to send a quantum state $|\psi\rangle$ to the receiver.
A priori condition necessary for this is that both sender and receiver share a 2-qubit entangled state and keep each bit (e-bit) with each other.
Then, sender entangles $|\psi\rangle$ with his e-bit and applies hadamard gate.
He measures both qubits and computational basis. The measurement results are in the form ``00", ``01", ``10" and ``11".
Based on these measurements, received applies various gates to e-bit which eventually becomes in the state of $|\psi\rangle$ as before.

\begin{algorithm}[!htbp]
\caption{sat-QFL - QFL for LEO Satellites}
\label{alg:hierarchical_satFL}
\begin{algorithmic}[1]
\State \textbf{Input:} $N$ rounds, $m$ main satellites $\{MainSat\}$, $s$ secondary satellites $\{SecSat\}$ for each $s_m \in \{MainSat\}$, 
$\{Geo\}$ Geo Stationary satellite,
training approaches: $\in \{\text{Sequential}, \text{Simultaneous}, \text{Asynchronous}\}$, accessTimes $(s_s, s_m) \mapsto \{(t_{\text{start}}, t_{\text{end}})\}$ (for Asynchronous)
\State \textbf{Output:} Global model weights $\theta_{avg}$
\State Initialize $\theta_{avg}$ randomly
\For{round $n = 1$ to $N$}
    \State Initialize empty dictionary $\mathcal{M}_{\text{weights}}$ for main satellite weights
    \For{each $s_m \in \{MainSat\}$ in \textbf{parallel}}
        \State Initialize empty $\mathcal{W}_m$.
        \If{training Sequentially}
            \State Set $\theta_s \gets \theta_{previousDevice}$
            \For{each $s_s \in \{SecSat\}$ sequentially}
                \State Train $s_s$ on private data to get $\theta_s$
                \State Send $\theta_s$ to the next satellite.
            \EndFor
        \ElsIf{training Simultaneously}
            \For{each $s_s \in \{SecSat\}$ in \textbf{parallel}}
                \State Train $s_s$ on private data to get $\theta_s$
                \State Append $\theta_s$ to $\mathcal{W}_m$
                \State Perform Averaging.
            \EndFor
        \ElsIf{training Asynchronously}
            \For{each $s_s \in \{SecSat\}$ in \textbf{parallel}}
                \State Train $d_s$ on private data to get $\theta_s$
                \If{Access True}
                    \State Append $\theta_s$ to $\mathcal{W}_m$
                \EndIf
            \EndFor
            \State Perform averaging.
        \EndIf
        \State Store $\theta_m$ in $\mathcal{M}_{\text{weights}}[s_m]$
        \State Further train with main satellites.
        \State Send each main satellites model to create global model. 
    \EndFor
\EndFor
\end{algorithmic}
\end{algorithm}

\begin{algorithm}[!htbp]
\caption{Secure sat}
\label{alg:secure_satFL}
\begin{algorithmic}[1]
\Statex \textbf{Input:} 
\Statex \quad Model parameters: $\theta, \varphi \in \{\mathbb{R}^d$, $\text{params}[:i]$; First $i$ items\}.
\Statex \textbf{Output:} 
\Statex \quad Encrypted parameters using QKD, Symmetric Encryption and transfer $\theta, \varphi$ via teleportation.

\State \textbf{Quantum Key Distribution (QKD)}
\State Generate private key $K \in \{0,1\}^n$ using QKD protocol (Algorithm \ref{alg:qkd}).
\State Define encryption function $E_K: \mathbb{R}^d \to \mathbb{R}^d$ and decryption $D_K: \mathbb{R}^d \to \mathbb{R}^d$ using $K$.
\State Encrypt parameters: $\theta' = E_K(\theta)$, $\varphi' = E_K(\varphi)$, where $E_K$ implements One-Time Pad ($x \mapsto x \oplus K$) or another scheme (e.g., Fernet).

\State \textbf{Quantum Teleportation}
\State Encode parameters parts $\theta, \varphi$ into quantum state $\ket{\psi} \in \mathbb{C}^2$ via unitary $U: \ket{\psi} = U(\theta, \varphi)\ket{0}$.
\State Execute teleportation protocol (Algorithm \ref{alg:teleportation}) to transmit $\ket{\psi}$.
\State Apply inverse unitary $U^\dagger: \ket{\psi} \mapsto \theta, \varphi$ to recover parameters.
\end{algorithmic}
\end{algorithm}

\begin{figure*}[ht]
    \centering
    \begin{subfigure}[b]{\linewidth}
        \includegraphics[width=\linewidth]{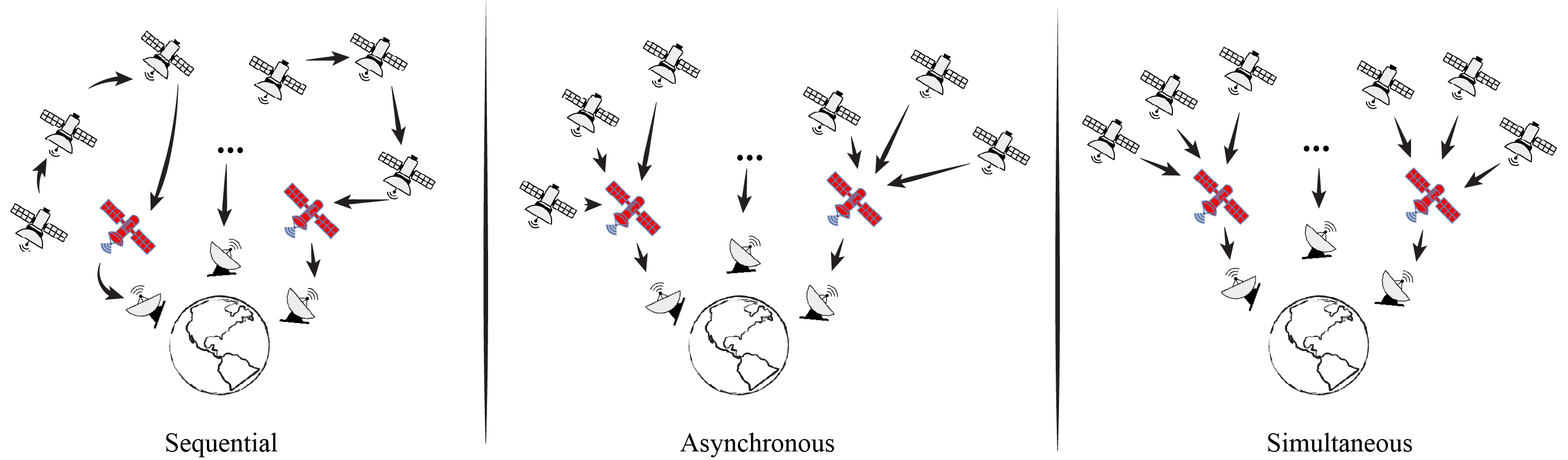}
        \caption{Representative sat-QFL frameworks at Secondary Satellites Level: Simultaneous, Sequential and Asynchronous}
        \label{fig:satQFL_framework_ico}
    \end{subfigure}
    \hfill
    \vspace{0.1cm}
    \begin{subfigure}[b]{\linewidth}
        \includegraphics[width=\linewidth]{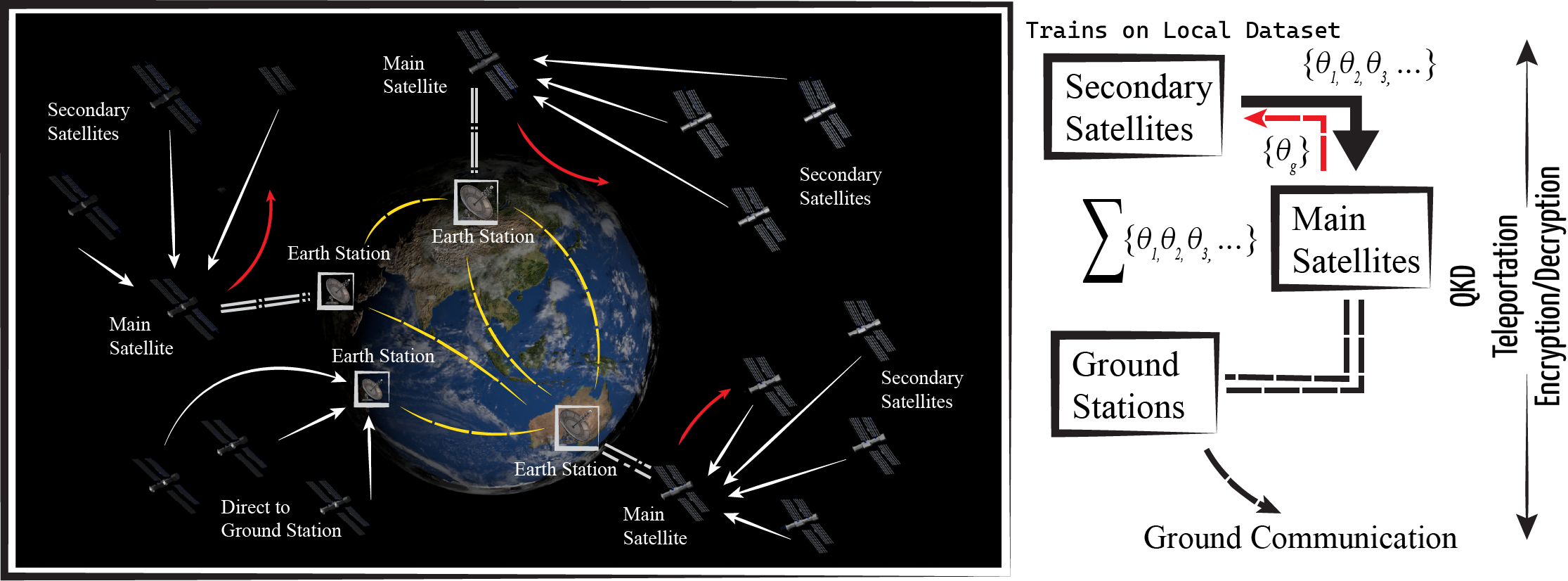}
        \caption{Overview of sat-QFL framework: LEO satellites collaboratively train global quantum model following specific topological network secured through quantum protocols.}
        \label{fig:satQFL_framework_full}
    \end{subfigure}
    \caption{Visualizations of the sat-QFL framework at different abstraction levels: Various topologies and approaches including sequential, asynchronous and simultaneous with security provide through QKD and Teleportation approaches integrated.}
    \label{fig:satQFL_framework_overall}
\end{figure*}

\section{Proposed Framework}
We propose a secure sat-QFL framework for LEO satellite constellations.
The proposed methods are presented below.

\subsection{sat-QFL}
LEO satellites are classified into  $m$ \textbf{main satellites} $\{MainSat\}_{m=1}^M$, capable of communicating with $p$ geostationary gateway stations $\{Geo\}_{m=1}^M$, and varying number of $s$ \textbf{secondary satellites} $\{SecSat\}_{s=1}^S$ ($s \in \mathcal{W} = \{0, 1, 2, ..\})$ for each $s_m \in \{MainSat\}$, which lack direct GEO connectivity but can connect to main satellites. The connectivity is modeled by a graph $G = (\{MainSat\} \cup \{SecSat\}, E)$, where edges $E$ indicate communication links, with some secondary satellites $s_s \in \{SecSat\}$ connecting to multiple main satellites $s_m \in \{MainSat\}$.

Our proposed framework operates at three levels.
\begin{enumerate}
    \item \textbf{At secondary (edge) satellite level:} We propose three training variants for secondary satellites $s_s \in \{SecSat\}$.
    \begin{enumerate}
        \item \textit{Sequential Training:} Each secondary satellite $s_s \in \{SecSat\}$ trains sequentially, updating model weights $\theta_s \in \mathbb{R}^d$ and passing them to the next satellite $s_{s+1}$. 
        The final satellite sends model weights $\theta_{s+1+..}$ to main satellite $S_m$.
        \item \textit{FedAveraging:} Each $s_s \in \{SecSat\}$ trains independently and sends weights $\theta_s \in \mathbb{R}^d$ to its accessible  $s_m \in \{MainSat\}$ via
        \begin{enumerate}
            \item \textit{Asynchronous Approach:} Secondary satellites transmit $\theta_s$ at only access times $\{(t_{\text{start}}, t_{\text{end}})\}$ to its main satellites.
            \item \textit{Simultaneous Approach:} All satellites transmit $\theta_s$ concurrently upon training completion to their main satellites.
        \end{enumerate}
    \end{enumerate}
    \item \textbf{At primary satellite level:} Each $s_m \in \{MainSat\}$ receives weights $\{\theta_s\}_{d_s \in \{SecSat\}}$. In the sequential approach, $s_m$ trains on the final $\theta_s$. In the federated approach, $s_m$ computes an averaged model $\theta_{avg} = \frac{1}{|\{MainSat\}|} \sum_{d_s \in \{SecSat\}} \theta_s$ and trains further on $\theta_{avg}$.
    \item \textbf{Gateway ground station level:} The ground station $G_m$ receives weights $\{\theta\}_{s_m}$ and performs global aggregation, computing $\theta_{avg} = \frac{1}{|\mathcal{M}|} \sum_{s_m \in \mathcal{M}} \theta_s$ for the global model.
\end{enumerate}

In Algorithm \ref{alg:hierarchical_satFL}, the set $\{MainSat\}$ of $m$ main satellites collaborates with their secondary satellites $\{SecSat\}$ to train a global model $\theta_{avg} \in \mathbb{R}^d$ in a hierarchical structure, defined by the connectivity graph $G$ and access times. As depicted in Figure \ref{fig:satQFL_framework_full}, main satellites $s_m$ connect to $Geo$, transmitting $\theta_m$ at specified times $t \in \mathbb{R}$, ensuring coordinated communication.

While the framework emphasizes coordination among ground stations, main, and 
secondary satellites, the secondary satellite level supports three training modes: 
sequential, asynchronous, and simultaneous. In \textit{sequential} mode, satellites 
train iteratively, updating $\theta_{\text{local}} \gets \theta_s$. 
In \textit{asynchronous} mode, training depends on access times $\{(t_{\text{start}}, 
t_{\text{end}})\}$ between $s_m$ and $s_s$. 
In \textit{simultaneous} mode, satellites train in parallel, computing $\theta_s$ concurrently.

\subsection{Secure sat-QFL}
To establish a quantum-resistant framework, 
we propose integrating QKD, Algorithm \ref{alg:qkd} and quantum teleportation (Algorithm \ref{alg:teleportation}). 
QKD, using protocols like BB84 \cite{bennett_quantum_2014}, generates a private key $K \in \{0,1\}^n$. 
Its security is grounded in the no-cloning theorem, which prohibits cloning an arbitrary state $\ket{\psi} \in \mathbb{C}^{2^n}$ 
via any unitary $C: \ket{\psi} \otimes \ket{0} \mapsto \ket{\psi} \otimes \ket{\psi}$, 
and Heisenberg’s uncertainty principle, ensuring non-commuting observables (e.g., $\sigma_x, \sigma_z$) detect 
eavesdropping through disturbances in the density matrix $\rho \in \mathcal{D}(\mathbb{C}^{2^n})$.

Encryption employs classical schemes: One-Time Pad (OTP), defined as $E_K: \mathbb{R}^d \to \mathbb{R}^d$, $E_K(x) = x \oplus K$, 
achieving information-theoretic security ($H(x | E_K(x), K) = H(x)$), or 
Fernet, using AES-128 for computational security. 
OTP requires key exchange, often via secure physical meetings \cite{williamsExplorationsQuantumComputing2011}. 
Post-quantum cryptographic schemes, such as lattice-based encryption, can be adopted for quantum resistance, defining $E_K$ over a module lattice.

The secure sat-QFL algorithm (Algorithm \ref{alg:secure_satFL}) combines 
QKD for secure key generation with encryption for transmitting 
model parameters.
We also propose to use Quantum teleportation to
encode parameters into a state $\ket{\psi} \in \mathbb{C}^{2^m}$, $m \geq \lceil \log_2 d \rceil$, 
via unitary $U: \ket{\psi} = U(\theta, \varphi) \ket{0}^{\otimes m}$, 
transmits $\ket{\psi}$ using an entangled state $\ket{\Phi^+} = \frac{1}{\sqrt{2}}(\ket{00} + \ket{11})$, 
and recovers $\theta, \varphi$ via $U^\dagger$. This ensures both security and privacy in parameter transmission.

The sat-QFL algorithm is a hierarchical federated learning framework for
LEO satellite networks, involving a global server $Geo$, main satellites $\{MainSat\}$, and secondary satellites $\{SecSat\}$ per main satellite $s_m$. 
For secondary satellites level, 
the objective is to minimize the global loss at secondary satellites level as 
\[
F(\theta_s^*) = \frac{1}{s} \sum_{s=1}^{\mathcal{S}} f_s (\theta_s)
\]
then, at main satellite level, 
\[
F(\theta_m^*) = \frac{1}{m} \sum_{m=1}^{\mathcal{M}} f_m(\theta_m).
\]

Thus, following the averaging approach, we have the following.

\[
F(\theta_g)  = \frac{1}{(m, s)} \sum_{m=1}^{\mathcal{M}} \sum_{s=1}^S f_s(\theta_s).
\]

Whereas, with sequential approach, we have

\[
F(\theta_s) = \frac{1}{s} \sum_{s=1}^{\mathcal{S}} f_{s-1}(\theta_{s-1}).
\]
With asynchronous approach, the main difference is that
federated aggregation is done only after completion of transfer of models from all secondary satellites.

The secure sat-QFL algorithm (Algorithm \ref{alg:secure_satFL}) enables the secure transmission of model parameters fully ($[\theta, \varphi, ...] \in \mathbb{R}^d$) or partially (\(\theta, \varphi) \in \mathbb{R}^d\),
using the QKD for encryption keys and quantum teleportation for parameter transmission. 
It combines classical symmetric encryption (e.g. one-time Pad or Fernet) with quantum protocols to ensure confidentiality. 

The algorithm's security relies on
quantum mechanics fidelity, on which the no-cloning theorem and entanglement monogamy prevent eavesdropping without detection.
Also, quantum devices (photon sources and detectors) are uncompromisable; side-channel attacks are out of scope.
Secure QKD Algorithm \ref{alg:qkd} (e.g., BB84, E91) provides unconditional security against quantum adversaries.
With entanglement distribution, preshared Bell pairs for teleportation (Algorithm \ref{alg:teleportation}) are secure.
Encoding Feasibility assures unitary \(U\) encodes \(\theta, \varphi\) into \(\ket{\psi} \in \mathbb{C}^2\), implying small \(d \leq n \text{ qubits}\) or multi-qubit systems for large models.

\section{Performance Evaluation}
In this section, we present our experimental results, explain the datasets and tools used, etc.
\subsection{Set Up}
\begin{figure}[!htb]
    \centering
    \begin{subfigure}[b]{0.8\columnwidth}
        \centering
        \includegraphics[width=\textwidth]{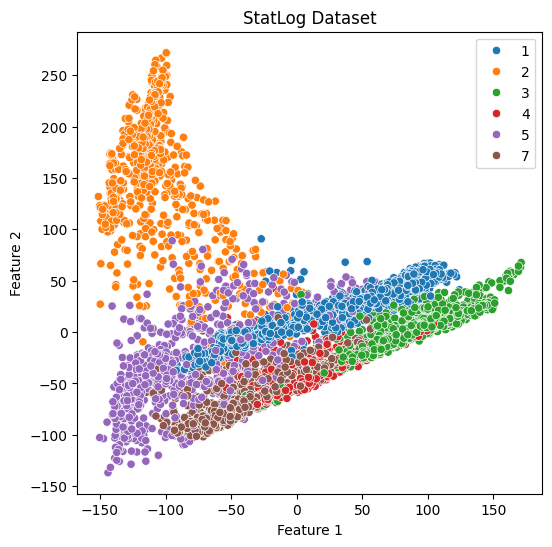}
        \caption{Statlog}
        \label{fig:datasets_statlog}
    \end{subfigure}
    \hfill
    \begin{subfigure}[b]{0.8\columnwidth}
        \centering
        \includegraphics[width=\textwidth]{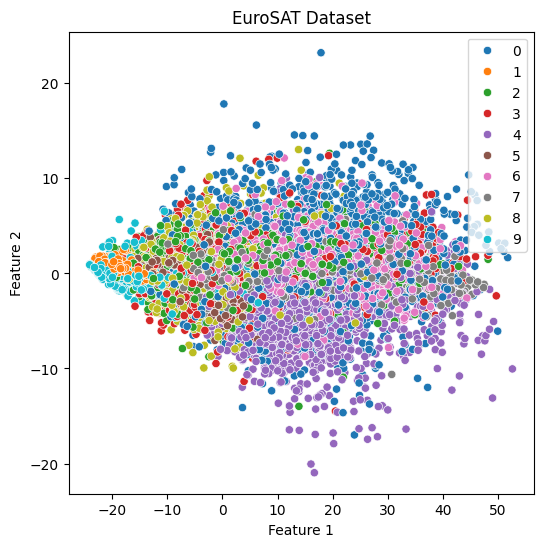}
        \caption{EuroSAT}
        \label{fig:datasets_eurosat}
    \end{subfigure}
    \caption{Dataset visualizations after applying dimension reductions}
    \label{fig:datasets_visualization}
\end{figure}

The first dataset that we used is the Statlog dataset (Landsat Satellite)
\cite{misc_statlog_(landsat_satellite)_146}.
It is a multivariate dataset that consists of multispectral values of pixels in $3$ x $3$ neighborhoods in satellite images.
In total, it has $6435$ samples with $36$ characteristics and $7$ labels from $1$ to $7$ where each number relates to red soil, cotton crop, gray soil, damp gray soil, soil with vegetation stubble, mixture class and very damp gray soil, respectively. 
Figure \ref{fig:datasets_statlog}, shows the Statlog datasets that are obtained after reducing the dimension of the feature.
In terms of data distribution among clients, $90\%$ of the dataset (distributed among satellites) is used for training, while the rest is used for testing (main server dataset).

Another dataset used is
EuroSAT dataset \cite{helber2017eurosat} \footnote{https://github.com/phelber/EuroSAT}, which is a land use and land cover classification dataset based on Sentinel-2 satellite images.
It contains $27,000$ images, each with a resolution of $64$ x $64$ pixels with $10$ land cover class labels (such as annual crop, forest, highway, pasture, etc.) collected from $34$ European countries.
Figure \ref{fig:datasets_eurosat} shows the EuroSAT dataset after dimension reduction using principal component analysis.

The tools used in the experimental analysis are Qiskit \footnote{\url{qiskit.org}}, Python, and Matlab \footnote{\url{https://www.mathworks.com/products/matlab.html}}.
Qiskit provides a specific library for quantum machine learning which introduces fundamental computational building blocks like quantum neural networks, variational quantum classifier, etc. which are used in this experimental analysis.
The initial handling of TLE data is conducted using MATLAB to facilitate its subsequent use in Python to simulate the proposed frameworks.

\textbf{Metrics.}
We have presented the experimental analysis of the framework based on the accuracy of the global model and local models. 
In addition, we conducted a performance comparison using objective values (loss) for both validation and training.
With regard to the state-of-the-art, we compare all of our models with the standard FedAvg algorithm.

\begin{figure}[!htbp]
\centering
    \begin{subfigure}{0.7\columnwidth}
        \centering
        \includegraphics[width=\columnwidth]{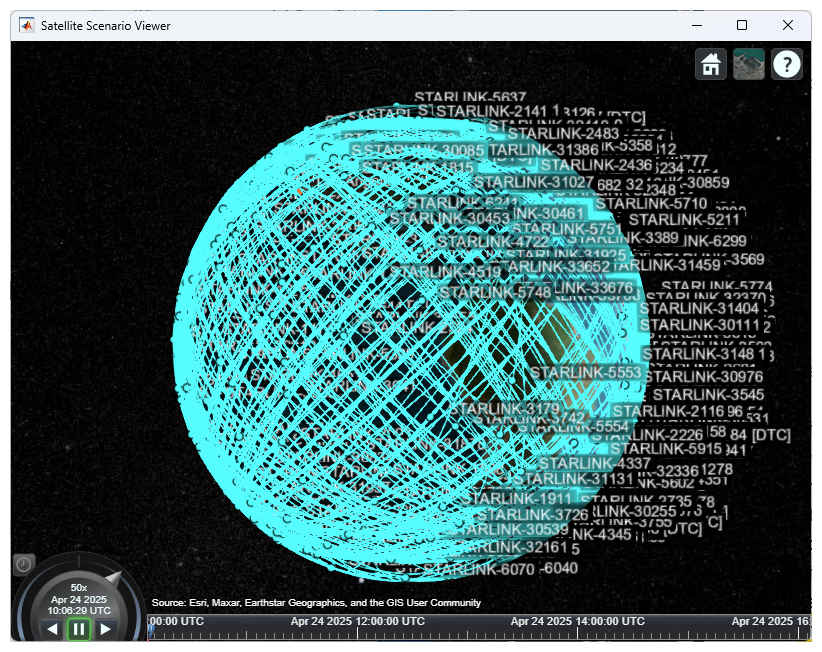}
        \caption{500 Satellites}
        \label{fig:starlink_500}
    \end{subfigure}
    \begin{subfigure}{0.7\columnwidth}
        \centering
    \includegraphics[width=\columnwidth]{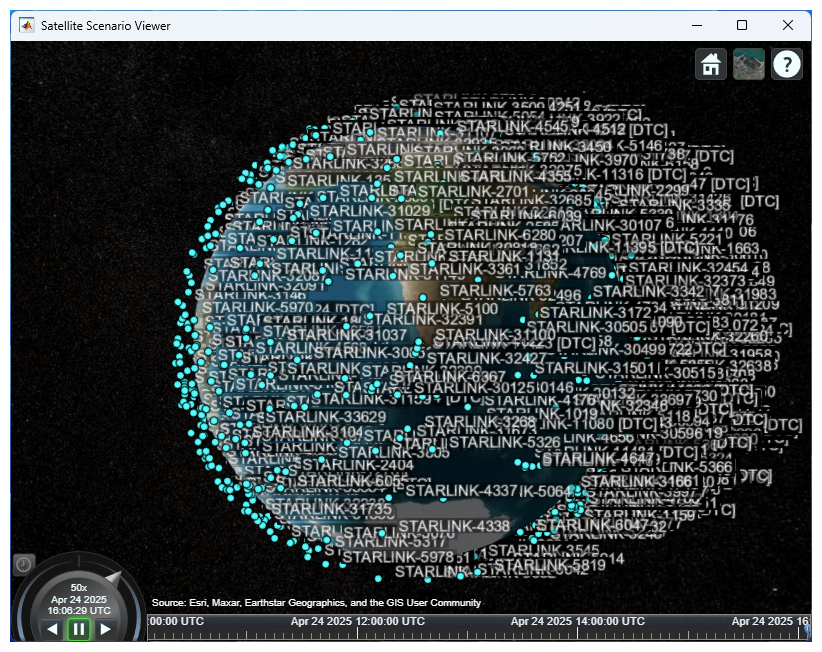}
        \caption{1000 Satellites}
        \label{fig:starlink_1000}
    \end{subfigure}
    \caption{Constellation simulation using MATLAB}
    \label{fig:constellation_simulation}
\end{figure}

\textbf{Satellite Scenario.}
In this work, we used Starlink TLE data to extract $50$ and $100$ satellites using MATLAB.
The figures \ref{fig:starlink_500} and 
\ref{fig:starlink_1000} shows the actual positioning of Starlink satellites according to the TLE document (Table \ref{tab:tle_data}, retrieved from \footnote{\url{https://celestrak.org/NORAD/elements/}}) of $500$ and $1000$ Starlink satellites.
We extract orbital parameters from the TLE document for $50$ satellites with simulation time window assumption from April $24, 2025, 10:06:29$ to $6$ hours later along with sample time i.e. satellite location time update every $30$ seconds.
We then attach canonical sensors (cameras) with maximum view angle to be $90$ degrees as shown in \ref{lst:scenario}.
We then define $10$ ground stations with their latitudes and longitudes such as Tokyo, Los Angeles, Madrid, 
Toronto etc.
At this stage, we evaluate the access from satellites to ground station by assessing each pair individually as well as via category, depending on whether there is direct access to the ground station or not.
We also compute satellite-to-satellite access and finally record the access intervals.
The results of the sample output are shown in Tables \ref{tab:satellite_mapping}.

\begin{minipage}{0.8\columnwidth}
\centering
\begin{lstlisting}[language=Matlab, label={lst:scenario}, basicstyle=\small]
% Define TLE file and scenario
numberofsatellites = 50;
tleFile = sprintf('starlink_%d.tle', numberofsatellites);
startTime = datetime(2025, 4, 24, 10, 6, 29);
stopTime = startTime + hours(6);
sampleTime = 30;
sc = satelliteScenario(startTime, stopTime, sampleTime);
\end{lstlisting}
\end{minipage}

\begin{table}[!htbp]
\centering
\caption{Main Satellites with Ground Stations and Non-Accessible Satellites (50 Satellites)}
\label{tab:satellite_mapping}
\resizebox{0.9\columnwidth}{!}{
\begin{tabular}{|l|l|p{6cm}|}
\hline
\textbf{Main Satellite} & \textbf{Ground Station} & \textbf{Non-Accessible Satellites} \\ \hline
STARLINK-4405 & Tokyo & STARLINK-11534, STARLINK-5398, STARLINK-11500, STARLINK-33604, STARLINK-32180, STARLINK-6068, STA... \\ \hline
STARLINK-30323 & Madrid & STARLINK-11534, STARLINK-31232, STARLINK-31820, STARLINK-31163, STARLINK-11573, STARLINK-5398, ST... \\ \hline
STARLINK-5594 & Santiago & STARLINK-11534, STARLINK-31820, STARLINK-31163, STARLINK-4280, STARLINK-11573, STARLINK-5398, STA... \\ \hline
STARLINK-1411 & Frankfurt & STARLINK-11534, STARLINK-31232, STARLINK-31820, STARLINK-31163, STARLINK-11573, STARLINK-32134, S... \\ \hline
STARLINK-2378 & Sydney & STARLINK-11534, STARLINK-4280, STARLINK-11573, STARLINK-32134, STARLINK-32360, STARLINK-11621, ST... \\ \hline
STARLINK-11243 & Bangalore & STARLINK-11534, STARLINK-4280, STARLINK-32933, STARLINK-31947, STARLINK-31063, STARLINK-3344, STA... \\ \hline
... \\
\hline
\end{tabular}
}
\end{table}

\begin{table}[!htbp]
\centering
\caption{TLE Data Sample for Starlink}
\label{tab:tle_data}
\resizebox{0.9\columnwidth}{!}{%
\begin{tabular}{|l|}
\hline
\textbf{STARLINK-1008} \\
\texttt{1 44714U 19074B   25112.58592294  .00005641  00000+0  39726-3 0  9991} \\
\texttt{2 44714  53.0538 188.1053 0001311  93.0175 267.0964 15.06401971300352} \\[0.2cm]
\midrule
\textbf{STARLINK-1010} \\
\texttt{1 44716U 19074D   25112.59326790 -.00012419  00000+0 -81623-3 0  9996} \\
\texttt{2 44716  53.0539 188.0720 0001737  85.8677 274.2511 15.06401699300336} \\[0.2cm]
\midrule
\textbf{STARLINK-1011} \\
\texttt{1 44717U 19074E   25113.71935023  .00027243  00000+0  18386-2 0  9991} \\
\texttt{2 44717  53.0549 203.0208 0001205  56.7504 303.3600 15.06443130300242} \\[0.2cm]
\midrule
\textbf{STARLINK-1012} \\
\texttt{1 44718U 19074F   25113.69169776 -.00005321  00000+0 -33866-3 0  9999} \\
\texttt{2 44718  53.0540 183.1402 0001426  87.0797 273.0355 15.06391628300518} \\[0.2cm]
\midrule
\textbf{STARLINK-1013} \\
\texttt{1 44719U 19074G   25112.43846484 -.00001219  00000+0 -62926-4 0  9991} \\
\texttt{2 44719  53.0540 188.7692 0001399 105.2021 254.9122 15.06400452301453} \\
\hline
$\cdots$ \\
\hline
\end{tabular}
}
\end{table}

\begin{figure*}[!htbp]
    \centering
    \includegraphics[width=\linewidth]{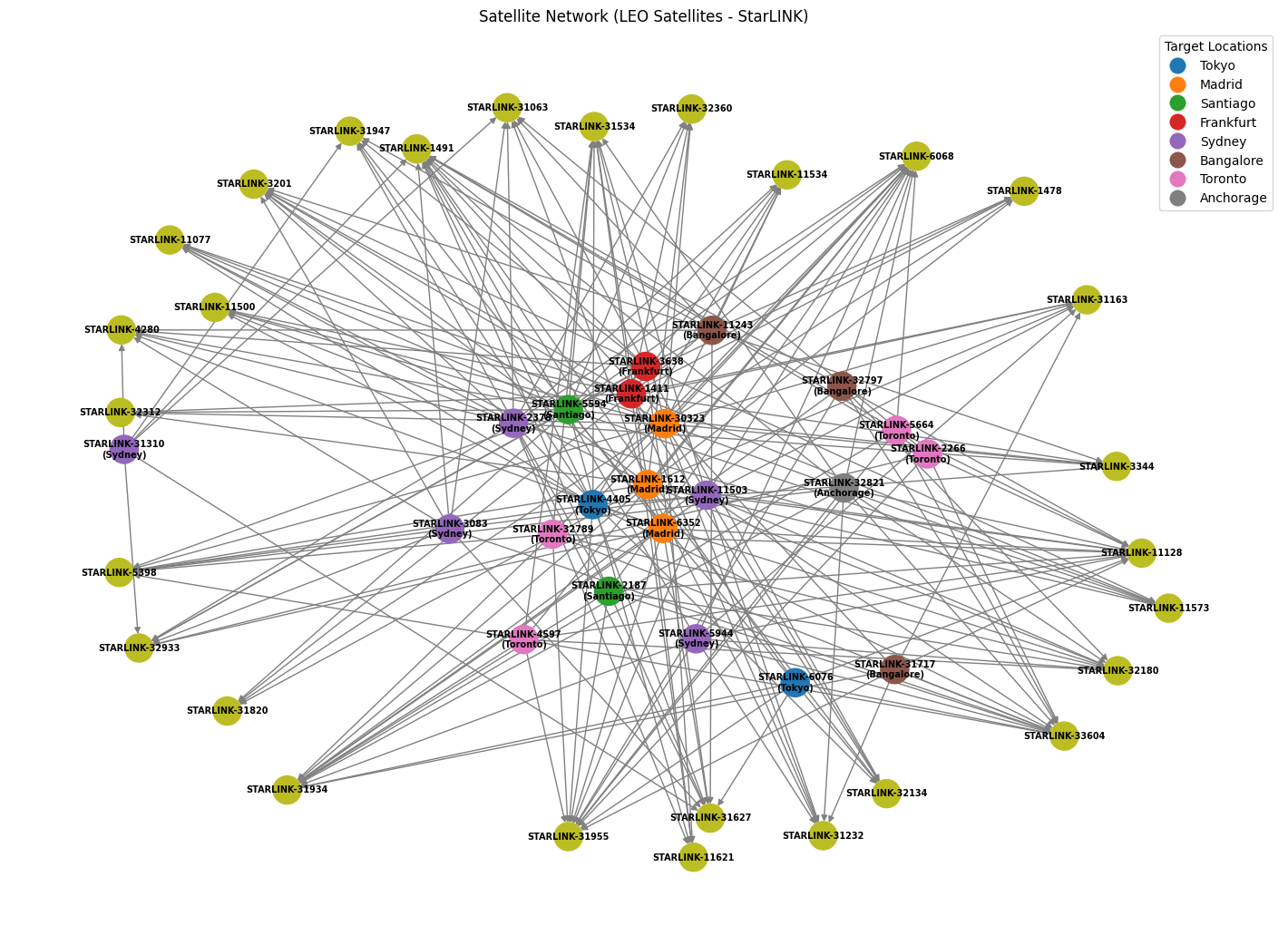}
    \caption{Network Diagram showing 50 Starlink LEO satellites, main satellites, secondary satellites and the target Geo locations.}
    \label{fig:network_50_3}
\end{figure*}

In Table \ref{tab:satellite_mapping}, we present the main satellite names, their ground stations, and the non-accessible satellites to ground stations.
Figure \ref{fig:network_50_3} shows the secondary satellites and the main satellites for the LEO Starlink satellites for 50 and link connection among multiple satellites.

\subsection{Results}
\begin{figure}[!h]
    \centering
    \begin{subfigure}[b]{0.45\columnwidth}
        \centering
       \includegraphics[width=\columnwidth]{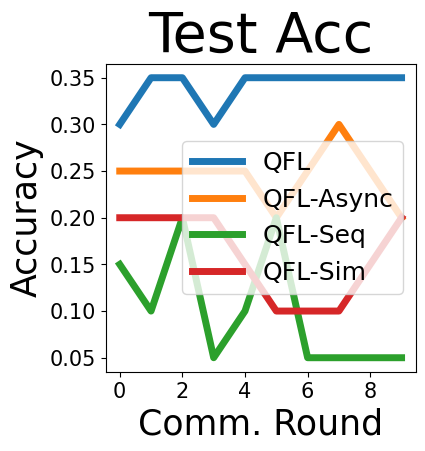}
    \caption{Test Acc; Statlog}
    \label{fig:server_test_acc_statlog}
    \end{subfigure}
    \centering  
  \begin{subfigure}[b]{0.45\columnwidth}
        \centering
      \includegraphics[width=\columnwidth]{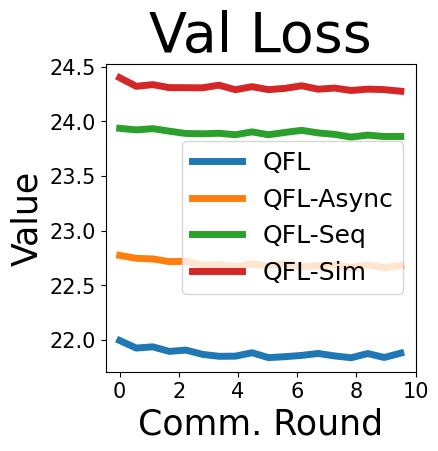}
    \caption{Val Loss; Statlog}
    \label{fig:server_obj_values_statlog}
    \end{subfigure}
    \begin{subfigure}[b]{0.45\columnwidth}
        \centering
       \includegraphics[width=\columnwidth]{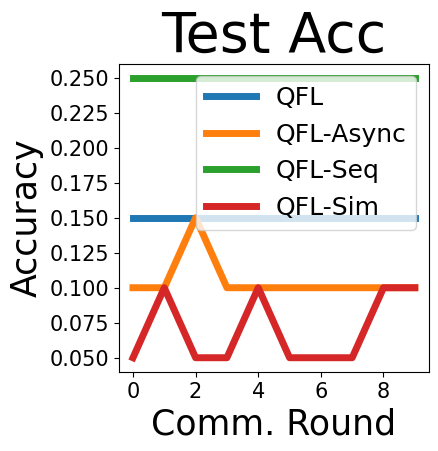}
    \caption{Test Acc; EuroSAT}
    \label{fig:server_test_acc_eurosat}
    \end{subfigure}
     \begin{subfigure}[b]{0.45\columnwidth}
        \centering
       \includegraphics[width=\columnwidth]{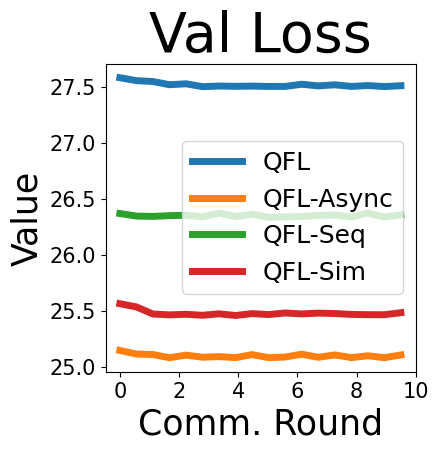}
    \caption{Val Loss; EuroSAT}
    \label{fig:server_obj_values_eurosat}
    \end{subfigure}
    \caption{Server Performance: EuroSAT and StatLog Dataset}
    \label{fig:server_performance}
\end{figure}

\textbf{Frameworks.}
In these basic results, we compare results between the frameworks presented, which are QFL, QFL-Async, QFL-Seq and QFL-Sim representing standard QFL, asynchronous QFL, sequential QFL and simultaneous QFL, respectively.
In terms of server test performance and validation loss as in Figures \ref{fig:server_test_acc_statlog}, \ref{fig:server_obj_values_statlog}, the default QFL performs better.
However, it should be clear that the standard QFL is not practical for satellite constellations and environments.
Whereas, with EuroSAT dataset as in Figures \ref{fig:server_test_acc_eurosat}, \ref{fig:server_obj_values_eurosat}, QFL-Seq performs better in terms of test accuracy, while QFL-Async performs best in terms of validation loss.

\begin{figure}[!h]
    \centering
    \begin{subfigure}[b]{0.45\columnwidth}
        \centering
       \includegraphics[width=\columnwidth]{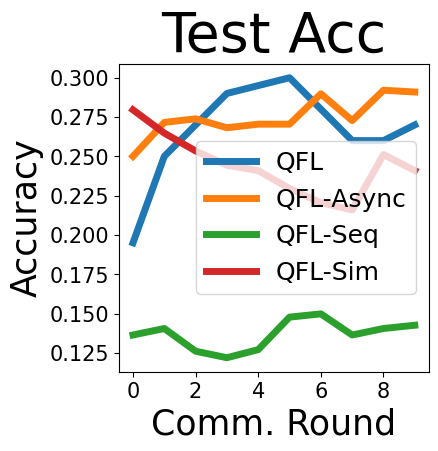}
    \caption{Test Acc; StatLog}
    \label{fig:devices_avg_test_acc_statlog}
    \end{subfigure}
    \begin{subfigure}[b]{0.45\columnwidth}
        \centering
       \includegraphics[width=\columnwidth]{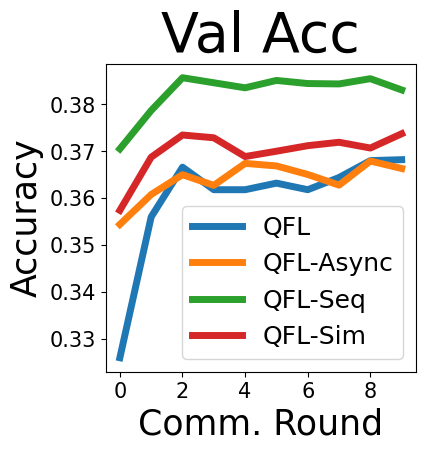}
    \caption{Val Acc; StatLog}
    \label{fig:devices_avg_val_acc_statlog}
    \end{subfigure}
     \begin{subfigure}[b]{0.45\columnwidth}
        \centering
       \includegraphics[width=\columnwidth]{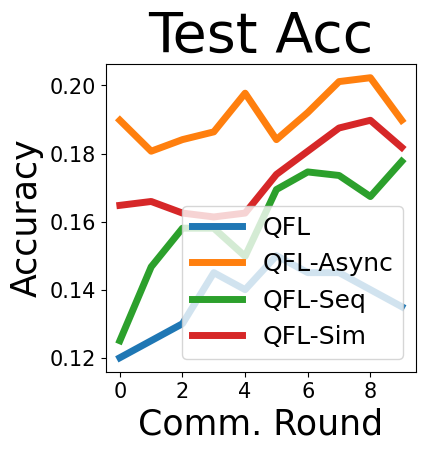}
    \caption{Test Acc; Eurosat}
    \label{fig:devices_avg_test_acc_eurosat}
    \end{subfigure}
     \begin{subfigure}[b]{0.45\columnwidth}
        \centering
       \includegraphics[width=\columnwidth]{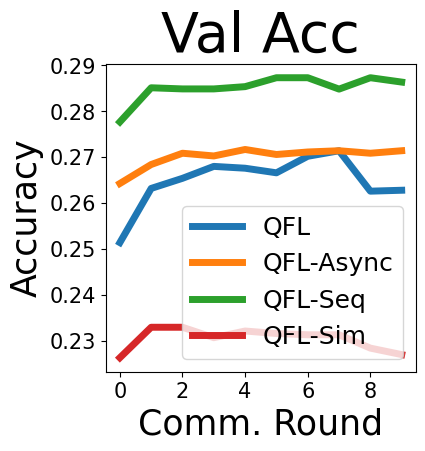}
    \caption{Val Acc; StatLog}
    \label{fig:devices_avg_val_acc_eurosat}
    \end{subfigure}
    \caption{Average Devices Performance: EuroSAT and StatLog Dataset}
    \label{fig:device_performance}
\end{figure}

In terms of device performance, with Statlog dataset as shown in Figure \ref{fig:devices_avg_test_acc_statlog} and \ref{fig:devices_avg_val_acc_eurosat}, test accuracy is better with QFL and QFL-Async while validation accuracy is better with QFL-Seq. 
Similarly, with the EuroSAT dataset as in Figures \ref{fig:devices_avg_test_acc_eurosat}, QFL-Async, QFL-sim and QFL-seq perform better in terms of test accuracy. Also in terms of validation accuracy, the performance is better with QFL-Seq as in Figure \ref{fig:devices_avg_val_acc_eurosat}.
Therefore, these findings demonstrate an enhanced effectiveness of the proposed frameworks compared to the standard QFL, especially when considering practical aspects.

\textbf{Teleportation.}
In this section, we compare the performance results between QFL and QFL with Teleportation.
However, it should be clear that teleportation itself is not supposed to impact the performance (accuracy) of the system, as teleportation is approach used to send quantum states. 
Thus, overall performance is somehow similar between the two.

\begin{figure}[!h]
    \centering
    \begin{subfigure}[b]{0.45\columnwidth}
        \centering
       \includegraphics[width=\columnwidth]{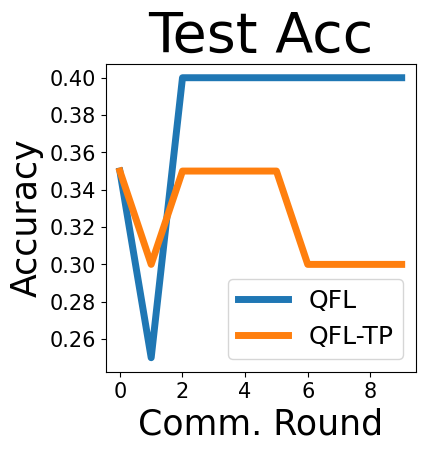}
    \caption{Test Acc; Statlog}
    \label{fig:server_test_acc_statlog_tp}
    \end{subfigure}
  \begin{subfigure}[b]{0.45\columnwidth}
        \centering
      \includegraphics[width=\columnwidth]{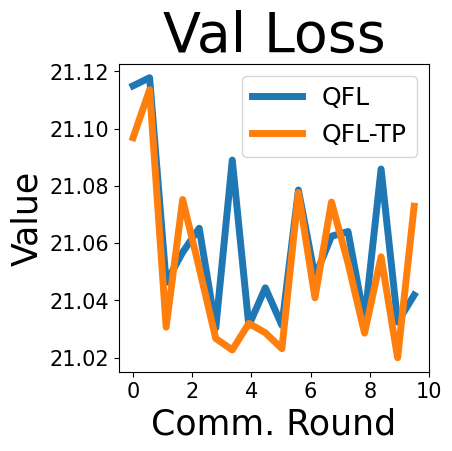}
    \caption{Val Loss; Statlog}
    \label{fig:server_obj_values_statlog_tp}
    \end{subfigure}
    \begin{subfigure}[b]{0.45\columnwidth}
        \centering
       \includegraphics[width=\columnwidth]{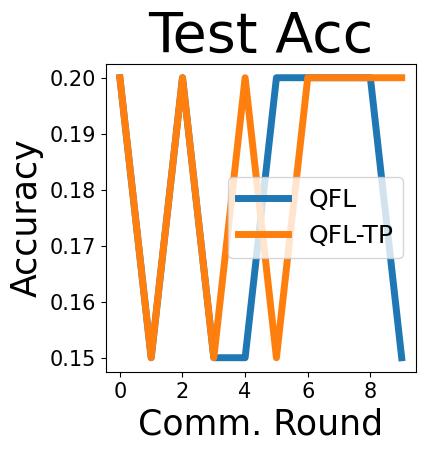}
    \caption{Test Acc; Eurosat}
    \label{fig:server_test_acc_eurosat_tp}
    \end{subfigure}
  \begin{subfigure}[b]{0.45\columnwidth}
        \centering
      \includegraphics[width=\columnwidth]{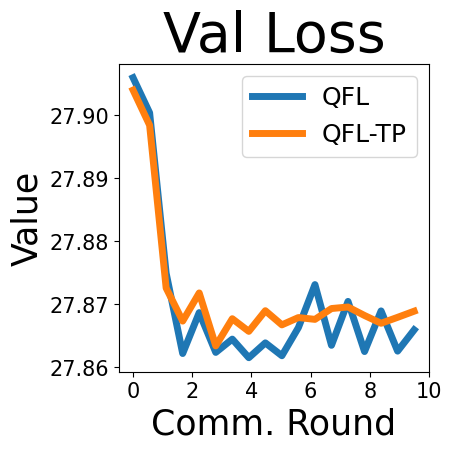}
    \caption{Val Loss; Eurosat}
    \label{fig:server_obj_values_eurosat_tp}
    \end{subfigure}
    \caption{Server Performance (Teleportation): Statlog and EuroSAT Dataset}
    \label{fig:server_performance_eurosat}
\end{figure}

With Statlog datasets, the test accuracy performance as in Figure \ref{fig:server_test_acc_statlog_tp} is better with QFL, but in terms of validation loss, the results are similar as in Figure \ref{fig:server_obj_values_statlog_tp} confirming little impact due to teleportation implementation.
However, with the EuroSAT dataset, the performance is almost similar with or without TP as in Figures \ref{fig:server_test_acc_eurosat_tp} for the accuracy of the server test and \ref{fig:server_obj_values_eurosat_tp} for the loss of server validation.

\begin{figure}[!h]
    \centering
    \begin{subfigure}[b]{0.45\columnwidth}
        \centering
       \includegraphics[width=\columnwidth]{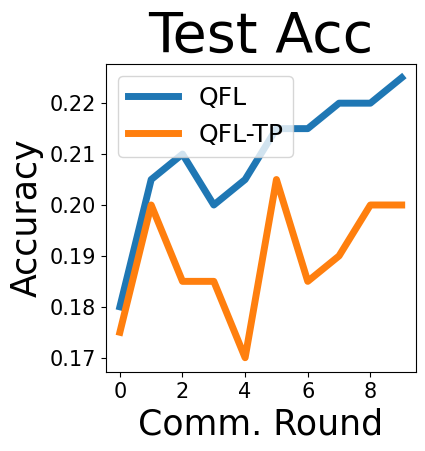}
    \caption{Test Acc; StatLog}
    \label{fig:devices_avg_test_acc_statlog_tp}
    \end{subfigure}
    \begin{subfigure}[b]{0.45\columnwidth}
        \centering
       \includegraphics[width=\columnwidth]{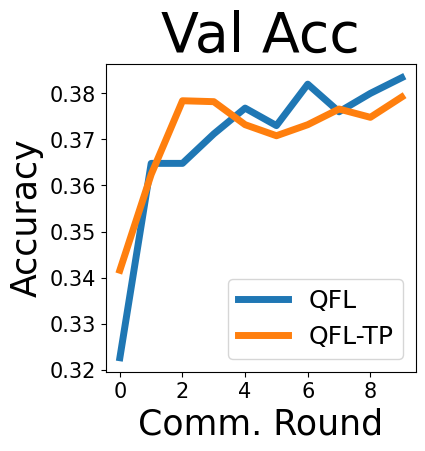}
    \caption{Val Acc; StatLog}
    \label{fig:devices_avg_val_acc_statlog_tp}
    \end{subfigure}
     \begin{subfigure}[b]{0.45\columnwidth}
        \centering
       \includegraphics[width=\columnwidth]{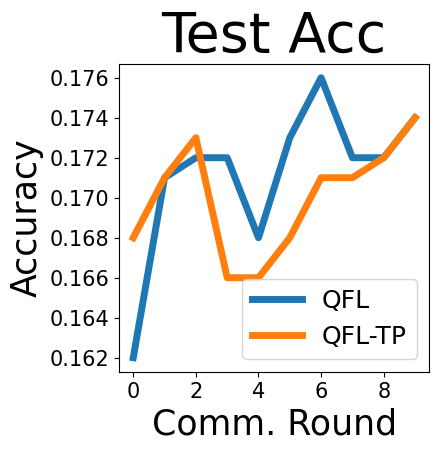}
    \caption{Test Acc; EuroSAT}
    \label{fig:devices_avg_test_acc_eurosat_tp}
    \end{subfigure}
    \begin{subfigure}[b]{0.45\columnwidth}
        \centering
       \includegraphics[width=\columnwidth]{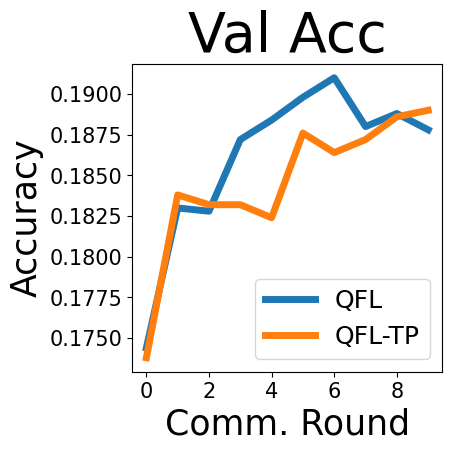}
    \caption{Val Acc; EuroSAT}
    \label{fig:devices_avg_val_acc_eurosat_tp}
    \end{subfigure}
    \caption{Devices Performance (Teleportation): EuroSAT and StatLog Dataset}
    \label{fig:device_performance_tp}
\end{figure}

The results are similar at the device level between QFL and QFL-TP in terms of test accuracy, validation accuracy, and training time as in Figure \ref{fig:device_performance_tp}.

\begin{figure}[!h]
    \centering
    \begin{subfigure}[b]{0.45\columnwidth}
        \centering
       \includegraphics[width=\columnwidth]{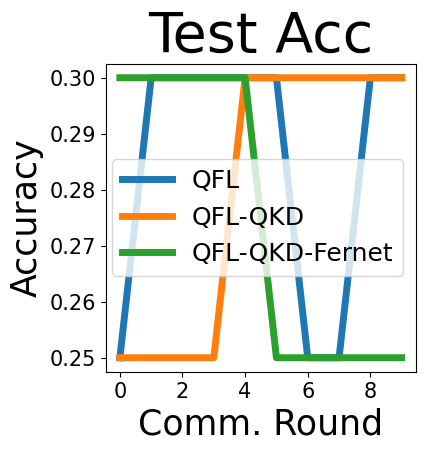}
    \caption{Test Acc; Statlog}
    \label{fig:server_test_acc_statlog_qkd}
    \end{subfigure}
  \begin{subfigure}[b]{0.45\columnwidth}
        \centering
      \includegraphics[width=\columnwidth]{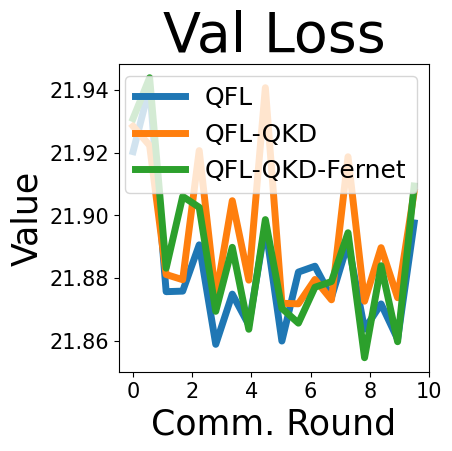}
    \caption{Val Loss; Statlog}
    \label{fig:server_obj_values_statlog_qkd}
    \end{subfigure}
     \begin{subfigure}[b]{0.45\columnwidth}
        \centering
       \includegraphics[width=\columnwidth]{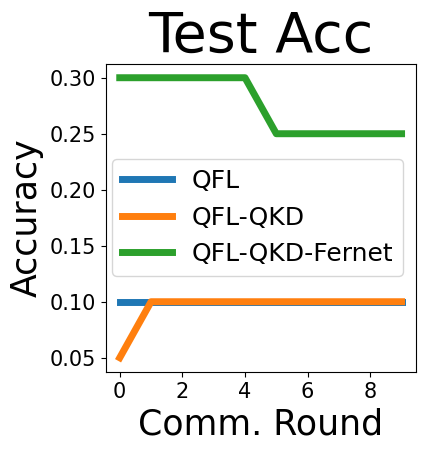}
    \caption{Test Acc; EuroSAT}
    \label{fig:server_test_acc_eurosat_qkd}
    \end{subfigure}
  \begin{subfigure}[b]{0.45\columnwidth}
        \centering
      \includegraphics[width=\columnwidth]{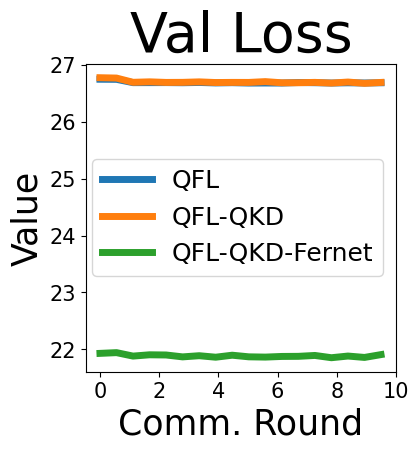}
    \caption{Val Loss; EuroSAT}
    \label{fig:server_obj_values_eurosat_qkd}
    \end{subfigure}
    \caption{Server Performance (QKD): StatLog and EuroSAT Dataset}
    \label{fig:server_performance_statlog_qkd}
\end{figure}

\begin{figure}[!h]
    \centering
    \begin{subfigure}[b]{0.45\columnwidth}
        \centering
       \includegraphics[width=\columnwidth]{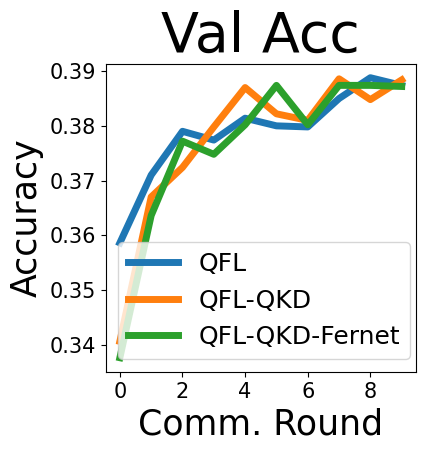}
    \caption{Val Acc; Statlog}
    \label{fig:devices_avg_val_acc_statlog_qkd}
    \end{subfigure}
    \begin{subfigure}[b]{0.45\columnwidth}
        \centering
       \includegraphics[width=\columnwidth]{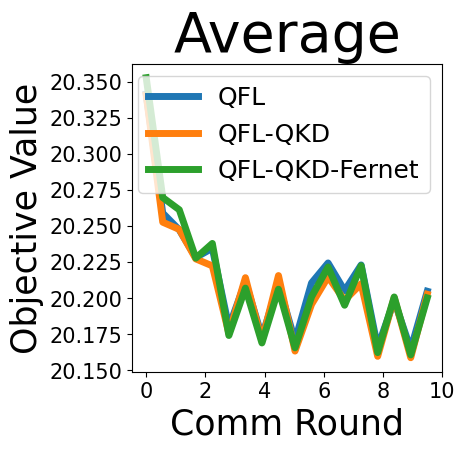}
    \caption{Val Loss; Statlog}
    \label{fig:devices_avg_obj_values_statlog_qkd}
    \end{subfigure}
  \begin{subfigure}[b]{0.45\columnwidth}
        \centering
      \includegraphics[width=\columnwidth]{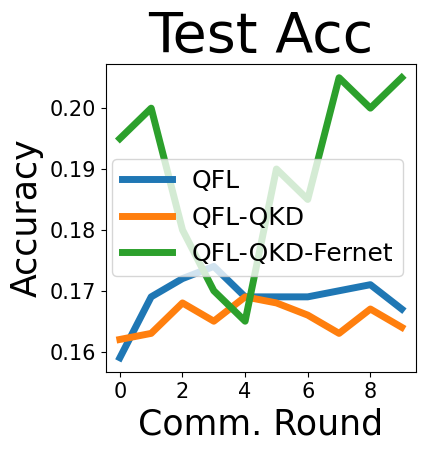}
    \caption{Test Acc; EuroSAT}
    \label{fig:devices_avg_test_acc_eurosat_qkd}
    \end{subfigure}
    \begin{subfigure}[b]{0.45\columnwidth}
        \centering
      \includegraphics[width=\columnwidth]{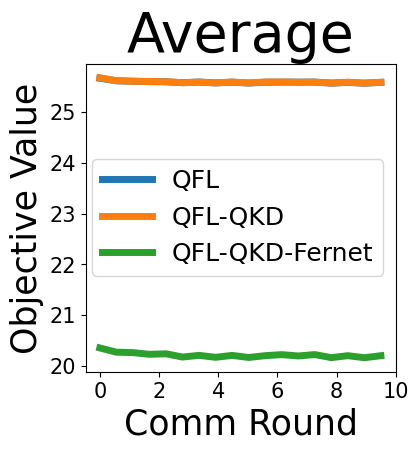}
    \caption{Val Loss; EuroSAT}
    \label{fig:devices_avg_obj_values_eurosat_qkd}
    \end{subfigure}
    
    \caption{Devices Performance (QKD): EuroSAT and StatLog Dataset}
    \label{fig:device_performance_eurosat_qkd}
\end{figure}

\textbf{QKD.}
In this section, we present experimental results in Figure \ref{fig:server_performance_statlog_qkd} with QFL, QFL-QKD and QFL-QKD-Fernet representing default QFL, QFL with QKD implementation, and QFL with QKD followed with fernet encryption and decryption implemented.
Again, as mentioned before, these implementations are not necessarily supposed to impact the overall performance of the system. However, the results vary as shown in the Figures \ref{fig:server_performance_statlog_qkd} and \ref{fig:device_performance_eurosat_qkd}.
In terms of server results, we observe varying results for Statlog dataset (Figures \ref{fig:server_test_acc_statlog_qkd}, \ref{fig:server_obj_values_statlog_qkd}).
 In Figures \ref{fig:server_test_acc_eurosat_qkd} and \ref{fig:server_obj_values_eurosat_qkd}, the results are better with QFL-QKD-Fernet than with QFL-QKD in terms of the EuroSAT dataset.
 
Similar results are observed in terms of device performance, with QFL-QKD-Fernet performing better (Figures \ref{fig:devices_avg_test_acc_eurosat_qkd}, \ref{fig:devices_avg_obj_values_eurosat_qkd})  with EuroSAT dataset. 
While with Statlog dataset, the results are similar validating no impact on the performance of the network with QKD integration as in Figures \ref{fig:devices_avg_val_acc_statlog_qkd} and \ref{fig:devices_avg_obj_values_statlog_qkd}.
However, the casualty of the results obtained just by implementing the QKD protocol with EuroSAT dataset needs to be studied further to understand in depth. 

\begin{figure}
    \centering
    \begin{subfigure}[b]{0.45\columnwidth}
        \centering
       \includegraphics[width=\linewidth]{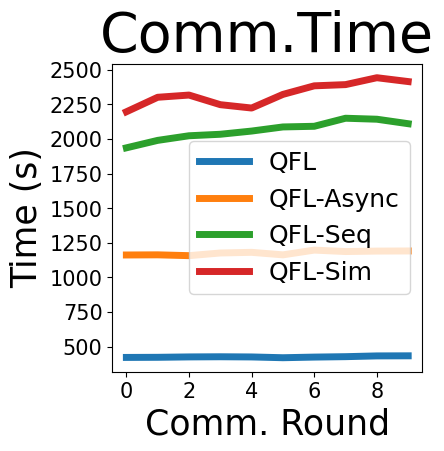}
    \caption{Frameworks}
    \label{fig:comm_time_statlog}
    \end{subfigure}
       \begin{subfigure}[b]{0.45\columnwidth}
        \centering
       \includegraphics[width=\linewidth]{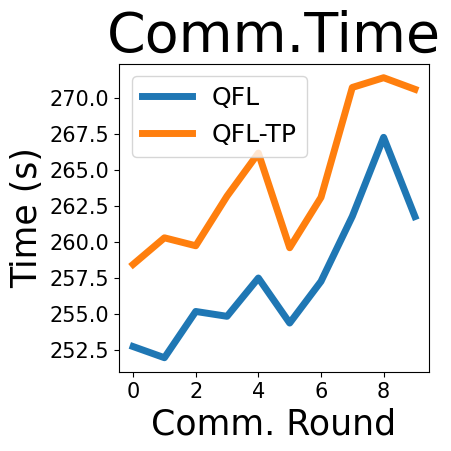}
    \caption{Teleportation}
    \label{fig:comm_time_statlog_tp}
    \end{subfigure}
       \begin{subfigure}[b]{0.45\columnwidth}
        \centering
       \includegraphics[width=\linewidth]{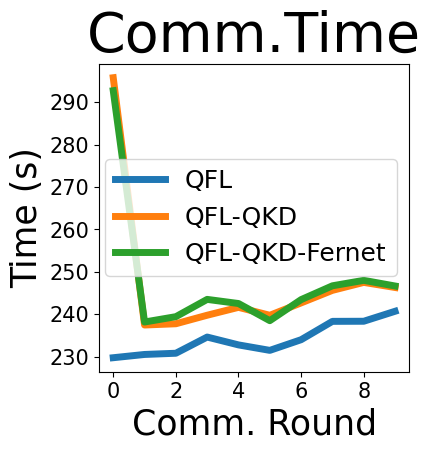}
    \caption{QKD}
    \label{fig:comm_time_statlog_qkd}
    \end{subfigure}
    \caption{Communication time; StatLog dataset}
    \label{fig:comm_time_datasets}
\end{figure}

In terms of communication rounds, QFL is better as in Figure \ref{fig:comm_time_datasets}.
Figure \ref{fig:comm_time_statlog} shows that QFL is faster than QFL-Async, QFL-Seq and QFL-Sim.
While QFL is also faster than QFL-TP as in Figure \ref{fig:comm_time_statlog_tp} which follows same in Figure \ref{fig:comm_time_statlog_qkd}.
This shows some communication overhead due to integration or implementation of the frameworks as presented in this work.
However, as pointed, standard QFL framework doesn't consider key satellite constellation factors such as access times, communication access etc., which are addressed in the proposed frameworks QFL-Async, QFL-Seq and QFL-Sim with added security with QFL-TP and QFL-QKD (also QFL-QKD-Fernet).
Thus, certain trade-off in terms of communication overhead is concluded.

The overall results are presented in tabular form in Table \ref{tab:basic_results}.
It shows additional results such as final accuracy at the end of all communication rounds, average results for both server and devices performance along with average communication overhead.
From the table, we can conclude that, in terms of communication time, there is tradeoff between practicality and performance.
Our proposed frameworks are practical considering they address critical satellite constellation issues.

\begin{algorithm}[!htbp]
\small
\setlength{\baselineskip}{0.9\baselineskip}
\caption{QKD}
\label{alg:qkd}
\begin{algorithmic}[1]
\State \textbf{Variables:} Message Chunks $C$, Key Size $n$
\Function{Prep}{$W$}
\State Initialize size of key, chunk of messages.
    \State Convert model weights $W$ into bytes.
    \State Decide the number of keys.
    \State Return keySize $n$, messageChunks $C$.
\EndFunction
\Function{GenerateRandom}{$n$}
    \State Initialize quantum register for circuit.
    \State Create $n$ number qubits in superpositions
    \State Execute circuit and extract counts 0s and 1s.
    \State Return rotation strings.
\EndFunction
\Function{GetMeasurement}{$C$}
    \State Generate random strings for sender and receiver.
    \State Choose sender and receiver rotation strings.
    \State Based on rotation strings, apply gates.
    \State Return measurement result.
\EndFunction
\Function{KeyGen}{$R_1, R_2, R$}
    \State Publicly shared rotation strings used.
    \State Keep same bits used.
    \State Create Key
\EndFunction
\end{algorithmic}
\end{algorithm}

\begin{algorithm}[!htbp]
\small
\caption{Quantum Teleportation}
\label{alg:teleportation}
\begin{algorithmic}[1]
\State \textbf{Input:} $\theta$, $\varphi$, Quantum Registers: $Q$ (secret), $A$ (Sender), $B$ (Receiver), Classical Register: $cr[3]$
\State Initialize quantum circuit $qc$ with $Q$, $A$, $B$, $cr$
    \State Entangle: Apply $H$ gate to $A$
    and $CNOT$ gate ($A$, $B$)
\State Create Secret: 
    Apply $U(\theta, \varphi, 0)$ gate to $Q$
\State Entangle Sender Secret:
    \State Apply $CNOT$ gate ($Q$, $A$)
   and Apply $H$ gate to $Q$
\State Measure:
    Measure $A$ into $cr[1]$
and $Q$ into $cr[0]$
\State Apply gates: X or Z gates.
    \If{$cr[1] = 1$}, Apply $X$ gate to $B$
    \EndIf
    \If{$cr[0] = 1$}, Apply $Z$ gate to $B$
    \EndIf
\State Receiver measurement:
  Measure $B$ into $cr[2]$
\end{algorithmic}
\end{algorithm}

\begin{table*}[h]
\centering
\caption{Basic framework results for sat-QFL}
\label{tab:basic_results}
\resizebox{\textwidth}{!}{
\begin{tabular}{llrrrrrrrrrrrrp{1cm}}
\toprule
Dataset & Model & \multicolumn{2}{c}{Server Val Acc} & \multicolumn{2}{c}{Server Test Acc} & \multicolumn{2}{c}{Server Val Loss} & \multicolumn{2}{c}{Devices Train Acc} & \multicolumn{2}{c}{Devices Test Acc} & \multicolumn{2}{c}{Devices Val Loss} & \multirow{2}{*}{Comm-Time (s)} \\
\cmidrule(lr){3-4} \cmidrule(lr){5-6} \cmidrule(lr){7-8} \cmidrule(lr){9-10} \cmidrule(lr){11-12} \cmidrule(lr){13-14}
 & & Avg & Final & Avg & Final & Avg & Final & Avg & Final & Avg & Final & Avg & Final & Avg\\
\midrule
\multirow{4}{*}{Statlog} & QFL & \textbf{0.268} & \textbf{0.28} & \textbf{0.34} & \textbf{0.35} & 21.88 & 21.88 & 0.36 & 0.37 & 0.27 & 0.27 & 20.78 & 20.76 & \textbf{426.46} \\
             & QFL-Async & 0.277 & \textbf{0.28} & 0.245 & 0.2 & 22.7 & 22.68 & 0.36 & 0.37 & \textbf{0.28} & \textbf{0.29} & nan & nan & 1177.27 \\
             & QFL-Seq & 0.245 & 0.25 & 0.1 & 0.05 & 23.89 & 23.86 & \textbf{0.38} & \textbf{0.38} & 0.14 & 0.14 & 20.62 & 20.60 & 2062.8 \\
             & QFL-Sim & 0.234 & 0.25 & 0.16 & 0.2 & 24.31 & 24.28 & 0.37 & 0.37 & 0.24 & 0.24 & \textbf{20.17} & \textbf{20.16} & 2324.88 \\
\midrule
\multirow{4}{*}{EuroSAT} & 
                QFL & 0.16 & 0.16 & 0.15 & 0.15 & 27.53 & 27.52 & 0.26 & 0.26 & 0.14 & 0.14 & 24.14 & 24.13 & \textbf{428.91} \\
             &  QFL-Async & 0.246 & 0.24 & 0.105 & 0.1 & 25.11 & 25.09 & 0.27 & 0.27 & 0.19 & 0.19 & nan & nan & 1184.13 \\
             & QFL-Seq & 0.16 & 0.16 & \textbf{0.25} & \textbf{0.25} & \textbf{26.35} & \textbf{26.35} & \textbf{0.29} & \textbf{0.29} & 0.16 & 0.18 & \textbf{24.06} & \textbf{24.04} & 2120.57 \\
             & QFL-Sim & \textbf{0.199} & \textbf{0.2} & 0.07 & 0.1 & 25.49 & 25.49 & 0.23 & 0.23 & \textbf{0.17} & \textbf{0.18} & 24.59 & 24.58 & 1995.06 \\         
\bottomrule
\end{tabular}
}
\end{table*}

\section{Conclusion}
In this work, we introduced the QFL framework, specifically tailored for satellite constellations, 
with a focus on LEO satellites. 
We examined various challenges associated with collaborative learning in these constellations. 
Our proposed framework, sat-QFL, demonstrates compatibility with LEO satellites in multiple respects, ensuring security in the quantum era. 
With both experimental and theoretical analysis, we highlighted limitations and challenges of LEO satellite constellations and validated our proposed frameworks, their practicality and suitability.

\end{document}